\documentclass[aip,twocolumn]{revtex4-1}
\usepackage{multirow}
\usepackage{graphicx}%
\usepackage{dcolumn}%
\usepackage{bm}%
\begin{document}


%\preprint{}

\title{ On the measurement of Adler angles in charged current single pion neutrino-nucleus interactions   }


\author{F.S\'anchez}
\email[]{fsanchez@ifae.es}
%\thanks{}
\affiliation{Institut de F\`{\i}sica d'Altes Energies (IFAE),  The Barcelona Institute of Science and Technology, Campus UAB. 01893 Bellaterra (Barcelona) Spain.}
%\affiliation{}


\date{\today}

\begin{abstract}
 
Uncertainties in modeling neutrino-nucleus interactions are a major contribution to  systematic errors in Long Base Line neutrino oscillation experiments.  
Accurate modeling of neutrino interactions requires additional experimental observables such as the Adler angles which carry information about the polarization of the $\Delta$ resonance and the interference with non-resonant single pion production. The Adler angles were measured with limited statistics in bubble chamber neutrino experiments as well as in electron-proton scattering experiments. We discuss the viability of measuring these angles in neutrino interactions with nuclei.

\end{abstract}

%\keywords{}

\maketitle

\section{Introduction}

The next generation of Long Base Line (LBL) accelerator neutrino oscillation experiments \cite{Ishida:2013kba,Adams:2013qkq} aims at the discovery of CP violation in the lepton sector and the determination of the neutrino mass hierarchy. Systematic errors are likely to limit the accuracy of these measurements because the energy region of this new generation of experiments, ranging from 0.5 to 10~GeV, is dominated by several poorly measured cross-section channels: charged and neutral current quasi-elastic scattering, single-pion production, multi-pion resonant production and collective nuclear responses such as short and long range nuclear correlations\cite{Nieves:2011pp}. Additional challenges are our incomplete understanding of the nuclear effects contributing to the cross-section and inaccuracies in reconstructing the neutrino energy. 

%The new generation of LBL near detectors  play an important role in controlling flux and cross.section systematics errors but some critical measurements will escape their experimental program. 

In neutrino oscillation experiments, accurate measurements of oscillation parameters demand that uncertainties arising from the errors on neutrino fluxes and from the neutrino interactions themselves be properly factorized. In turn, this requires the correct modeling of neutrino-nucleus cross sections channels. 
Neutrino cross-section knowledge suffers from three levels of uncertainties: i) Cross-sections at the nucleon level are not perfectly known. Vector form factors are derived from electron scattering, but axial and pseudoscalar form factors are assumed to be dipolar and constrained by the PCAC hypothesis. ii) Cross-sections are modified by effects due to the nuclear medium through short- and long-range correlations and by uncertainties in nucleon kinematics inside the nucleus. iii) The particles produced in the primary interaction cross the high-density nuclear medium which alters the particle composition of the event. Experimentally, the picture is confused even further by the typically broad neutrino energy spectrum and by beam flux uncertainties. 
%The modelling of neutrino-nucleus cross-sections is a critical input for the neutrino oscillation program. 
During the last decade efforts were focused on the description and measurement of quasi-elastic scattering; however we now also need to accurately model interactions occurring at higher energies such as single pion production. These  data are sparse and contradictory even at the nucleon level \cite{Wilkinson:2014yfa}. In addition, current and future  measurements can be performed only on nuclei. 
%We explore in this study the capability of using angular observables to improve our knowledge of this cross-section using neutrino nucleus cross-sections. 

In this paper, we explore the possibility of measuring Adler angles~\cite{Adler:1968tw}  in neutrino charged-current pion production on nuclei. A full description of the theoretical implications of Adler angles measurements can be found in the references \cite{Hernandez:2007qq}.

Experimental results from electron scattering on hydrogen have been published by the CLAS  \cite{Park:2007tn, Egiyan:2006ks} collaboration and previously by earlier  experiments \cite{Breuker:1977vy}. These results were obtained on  hydrogen targets, thereby  avoiding initial and final state nuclear interactions  and with the advantage of knowing the initial electron kinematics.  
Here, we discuss the possibility of measuring these angles in modern neutrino experiments, which are typically performed in broad-band beams and using heavy nuclear targets. In these experiments the neutrino energy must be reconstructed from the data, the initial target nucleon is not at rest and the final state particles undergo nuclear re-scattering. 

This work is based on the NEUT \cite{Hayato:2009zz} Monte Carlo model, described in the following section. The angular observables are illustrated next, followed by Monte Carlo results and conclusions. 

\section{ Monte Carlo Model }

Neutrino interactions are simulated with the NEUT \cite{Hayato:2009zz} program libraries, which include neutral current (NC) and charged  current (CC) processes of elastic and quasi-elastic scattering, meson exchange currents, single meson production, single gamma production, coherent pion production and non-resonant inelastic scattering. 

NEUT uses the Rein-Sehgal\cite{Rein:1980wg}  model to simulate neutrino-induced single pion production and an {\it ad-hoc} model for multiple pion production up to a hadronic invariant mass of 2.0~GeV/c$^2$. NEUT also includes a contribution from non-resonant pion production and the  $\Delta$ polarization values measured in deuterium \cite{Radecky:1981fn}. 

Final State Interactions (FSI) of hadrons taking place within the nuclear medium are also simulated using a microscopic cascade model. In the case of final state pions, the considered processes are inelastic scattering, pion absorption and charge exchange. The simulated nucleon interactions are elastic scattering as well as single and double $\Delta$ production. FSI interactions alter both the multiplicity of pions in the final state as well as the kinematics of the pions. 

%The results from the paper will depend strongly on the exact implementation of the FSI.  

For the current evaluation, events are generated with a T2K neutrino spectrum \cite{Abe:2012av}. The simulation includes all channels present in the NEUT  Monte Carlo.  Most current  experiments measure neutrino cross-sections on carbon-based detectors, such as plastic scintillator (Polystyrene).  To avoid confusion between hydrogen and carbon targets, we decided to study interactions on carbon nuclei. In the case of Polystyrene there will be an additional $\approx$15\% of interactions occurring with a free proton, in which the  neutrino energy and the Adler angles will be well-determined, except for detector effects. 

We consider both the $\Delta^{++}$ and the $\Delta^{+}$ decaying to a $\pi^+$ because they are indistinguishable at the experimental level for most of the final state nucleon momenta. 

\section{Adler Angles}
\label{Sec:Adler}

\begin{figure*}
\includegraphics[width=15cm]{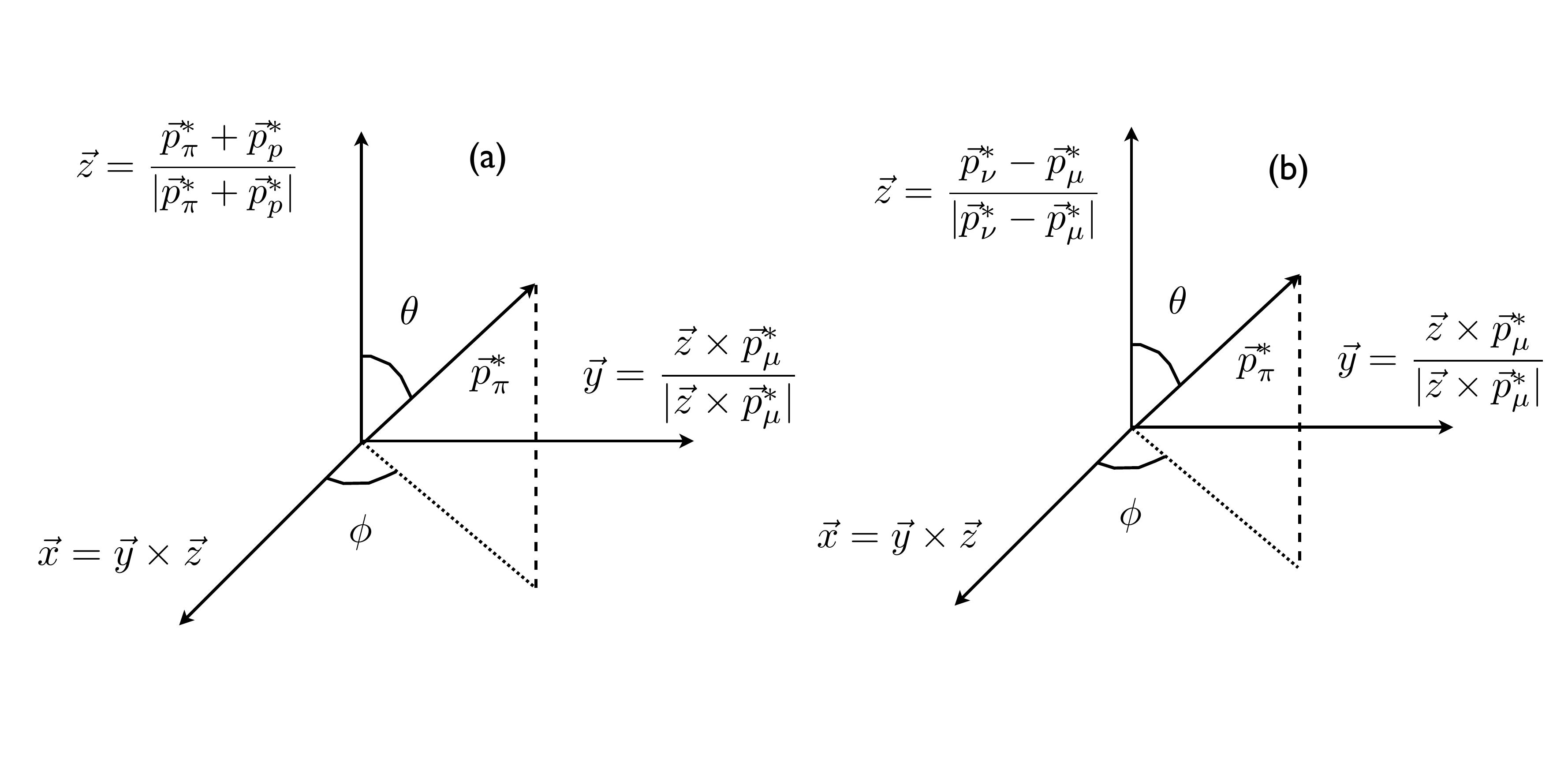}
\caption{\label{Fig:AdlerReference} Definition of the Adler Angles  at the nucleon (true) level (a) and the nuclear level (b). The momenta of the particles are defined in the $\vec q  = \vec p_{\nu}-\vec p_{\mu}$ rest frame.}
\end{figure*}

 The Adler reference system~\cite{Adler:1968tw} describes the  p$\pi^+$ final state in the p$\pi^+$ reference system. The two angles are defined as in Fig.\ref{Fig:AdlerReference}, where $\phi$ and $\theta$ are sensitive to the transverse and longitudinal polarization of the p$\pi^+$ system for interactions mediated by a $\Delta^{++}$, $\Delta^{+}$ and for non-resonant contributions.  The two angles are properly defined at the nucleon interaction level but they are altered by the Final State Interactions and theFermi momentum of the target nucleon.

%We distinguish between the nucleon (or true) level and the nucleus level where the angles should be redefined. 

\subsection { Adler angles at the level of the nucleus} 

\begin{figure}
\includegraphics[width=8.5cm]{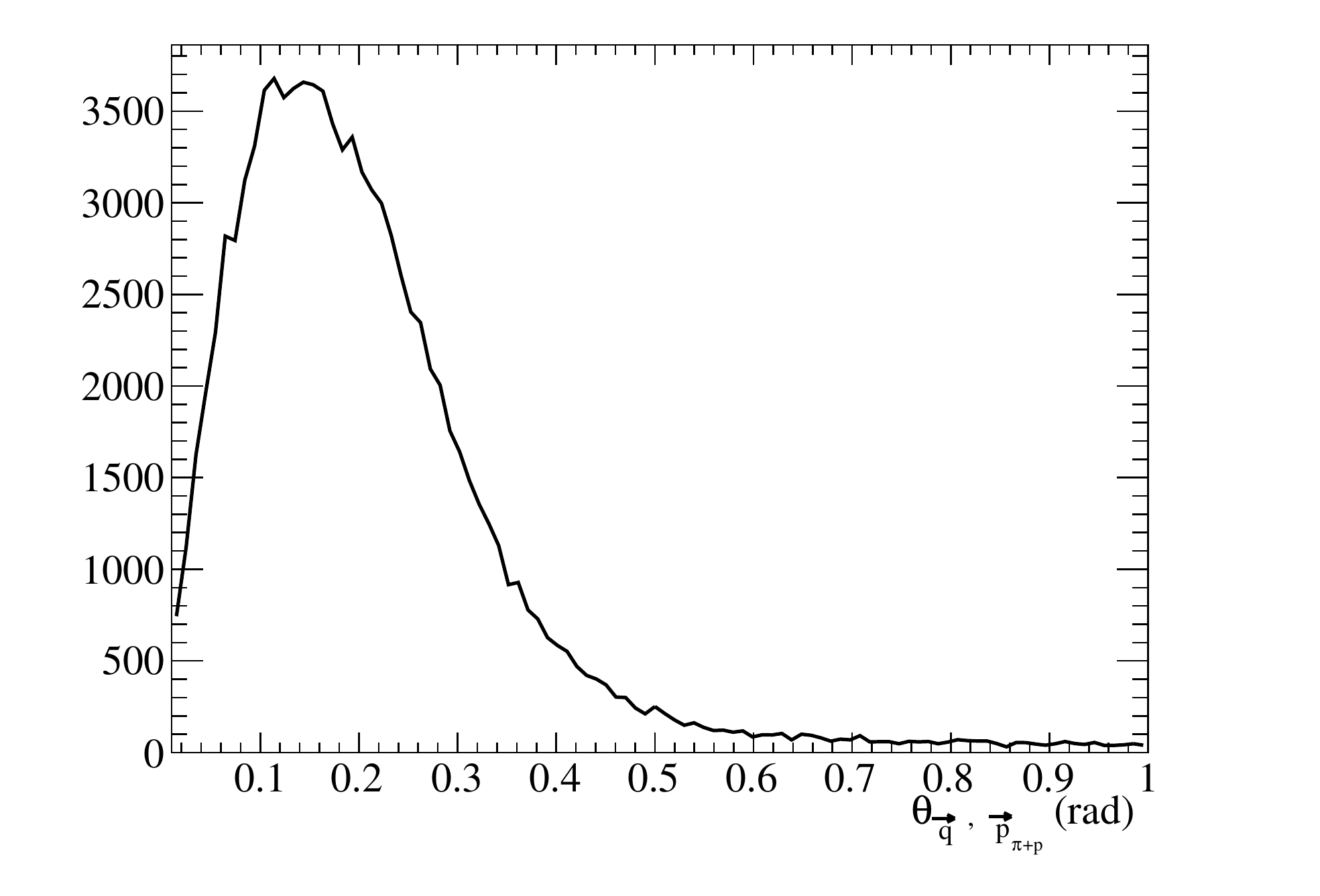}
\caption{\label{Fig:QdeltaAngle} Angle between the $\vec p_{p} + \vec  p_{\pi}$  system and the $\vec q$ approximation at the level of the nucleus.}
\end{figure}

  Modern experiments detect neutrino interactions on targets consisting of relatively heavy nuclei (carbon, oxygen, iron, argon); therefore the definition of the Adler angles needs to be modified. The first modification is mandated by the fact that normally the proton is not detected. In this case, the p$\pi^+$ reference system needs to be redefined based on detector observables, namely the lepton and the $\pi^+$.  In addition, we reconstruct the neutrino energy assuming that the target nucleon is at rest, thereby ignoring its intrinsic Fermi momentum, and assuming that the neutrino direction is known. In this scenario, energy-momentum conservation allows to estimate the neutrino energy as : 

$$E_{\nu} =   { m_p^2 - A_p^2  + | \vec p_{\mu} + \vec p_{\pi} |^2  \over 2 ( A_p +  \vec d_{\nu} \cdot ( \vec p_{\mu} + \vec p_{\pi} )  ) }  $$
$$ A_p =  m_p - E_{bind} - E_{\mu} - E_{\pi} $$ 

 where $(E_{\mu}, \vec p_{\mu})$ and $(E_{\pi}, \vec p_{\pi})$ are the four-momenta of the muon and the pion, $\vec d_{\nu}$ is the neutrino direction, $E_{bind}$ is the target nucleon binding energy ($\approx$~25~MeV in NEUT for a carbon target) and $m_p$ is the free proton mass. The target nucleon momentum cannot be inferred when the outgoing proton is not detected. The uncertainty introduced by this approximation will be discussed later. This definition of the neutrino energy and the assumption that the  target nucleon is at rest allow us to calculate the invariant mass of the p-$\pi$ system and to estimate the values of the observables used in deuterium experiments \cite{ Barish:1978pj, Radecky:1981fn}.
We approximate the direction of the final p-$\pi^+$ system by the momentum transfer to the nucleus ($\vec q = \vec p_{\nu}-\vec p_{\mu}$).  
%This approximation is forced by the fact that the target nucleon momentum cannot be obtained in the absence of proton detection. 
The angle between the true p-$\pi^+$ system and the estimated $\vec q$ is shown in Figure \ref{Fig:QdeltaAngle}. Under the aforementioned assumptions this approximation causes a bias of about 0.2~rad. 

The same observables can be reconstructed for the cases of $\Delta^- \rightarrow n \pi^-$ and $\Delta^0 \rightarrow p \pi^-$  to measure the Delta polarization in anti-neutrino nucleus interactions. 

\subsection { Fermi momentum versus Final State Interactions } 

To evaluate the relative contributions to the Adler angles of the Fermi momentum and the FSI, we compute the Adler angles under three assumptions:  
{\bf i) true:} we estimate the parameters using the full kinematic information at the level of the nucleon. These results are experimentally measurable only with a hydrogen target. 
{\bf ii) pre-FSI:} we use the true kinematics of the pion at the level of the nucleon  but we ignore the target nucleons momentum. In this case the effect of the Fermi momentum is taken into account but the FSIs are ignored. 
{\bf iii) post-FSI: } we use the information of the pion leaving the nucleus and ignore the kinematic information of the target nucleon. These are the actual experimental  observables and they contain the effect of both the Fermi momentum and of the FSI. 

\section{ Monte Carlo predictions }
\label{Sec:Results}

\subsection{ Selection of events and their categories }

Events are identified from the interactions at the nucleon level and from the multiplicity of particles leaving the nucleus. For the first criterion we rely on the Monte Carlo code to tag single pion production events, produced resonantly or not according to the model of Rein and Sehgal \cite{Rein:1980wg}. In order to assign the multiplicity of the pion final state we look for the number of  $\pi^+$, $\pi^-$, $\pi^0$ and $e^{\pm}$ emitted by the nucleus, after the FSI. We define a one-$\pi^+$ topology when one and only one $\pi^+$ is emitted by the nucleus and no other particle from the above list is present. We do not count emitted protons, neutron and nuclear gammas because they are often produced but in current experiments they are detected with low efficiency.  Nuclear de-excitation gammas play a negligible role in the description of the neutrino-nucleus interactions discussed here.

The one-$\pi^+$ events are then divided into three categories according to the true type of nucleon interaction: i) single pion production, the signal we want to study ii) deep inelastic scattering iii) other processes: for example, $\eta$ and kaon production.  In what follows the last two categories are considered background.  The fraction of true charged-current one-$\pi^+$ reactions in which a single $\pi^+$ emerges from the nucleus is $\sim$~75\%. In $\sim$~43\% of them the pion momentum is altered by FSI. The background is estimated to be $\sim$~18\% of single $\pi^+$ events leaving the nucleus. These numbers depend on the error on the actual neutrino flux and on the MC models; they only have indicative value. 

\begin{figure}

\includegraphics[width=8cm]{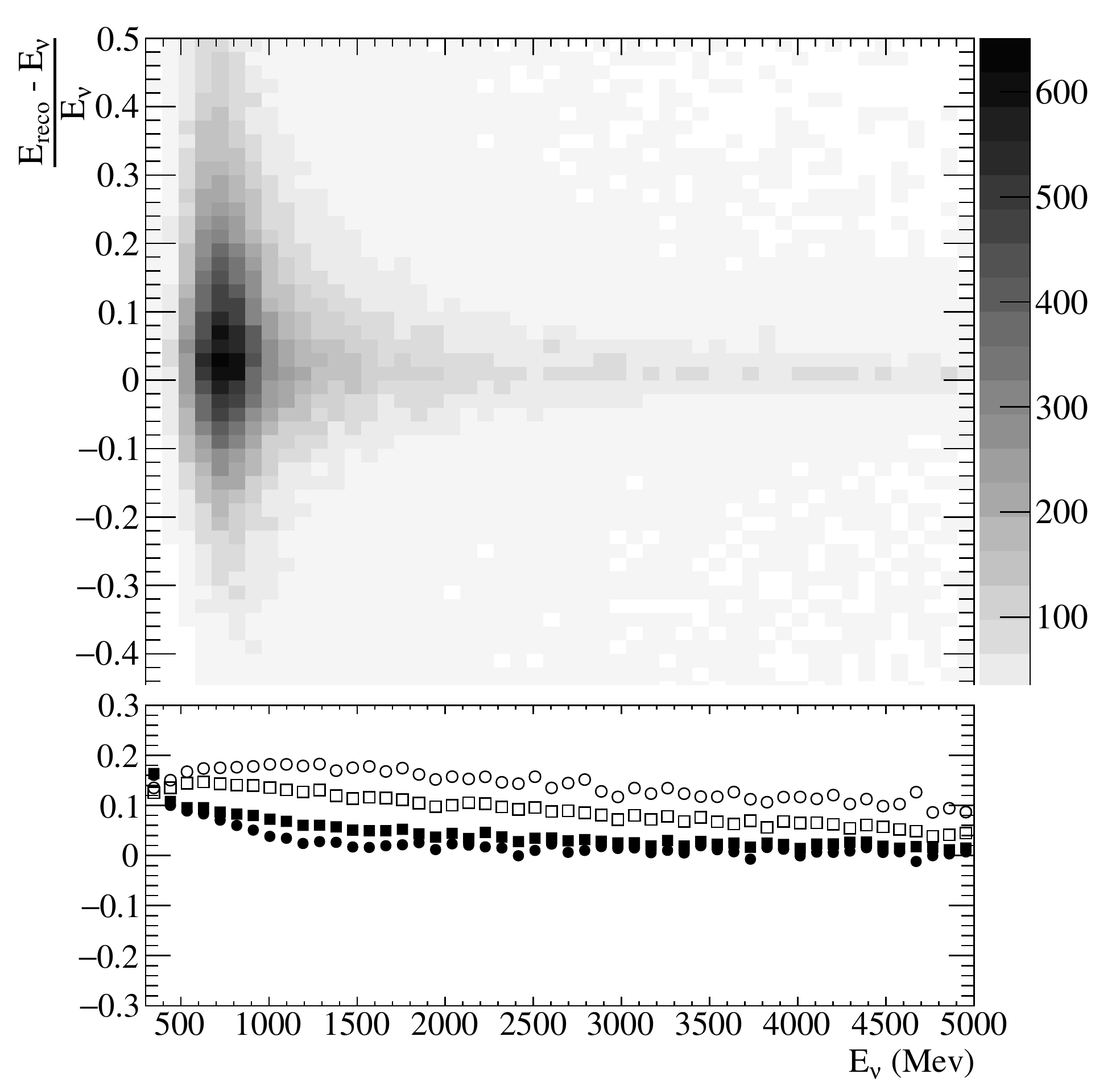}
\caption{\label{Fig:dEnuEnu}  Relative spread of the reconstructed neutrino energy $ ( E_{reco} - E_{\nu} ) / E_{\nu} $ as a function of the true neutrino energy. The black circles (squares) represent the mean value of the relative difference post(pre)-FSI.  The empty circles (squares) show the RMS of the reconstructed neutrino energy relative error post(pre)-FSI. }
\end{figure}

\subsection{ Neutrino energy and hadronic invariant mass reconstruction }

	The relative spread of the reconstructed neutrino energy is shown in Figure \ref{Fig:dEnuEnu}  as a function of the neutrino energy. At low energy, due to the relatively large contribution of the target nucleon Fermi momentum, the bias is large, but it decreases from 10\% to almost zero at higher neutrino energies due the reduced contribution of the target´s Fermi momentum to the total interaction energy. The RMS error is also larger at low energies where it reaches 20\% while it decreases to 10\% for high neutrino energies.  Opposite to other observables, the reconstructed pre-FSI energy is biased towards higher values of energy than the post-FSI reconstruction.  The FSI compensates, through the pion energy loss inside the nucleus, the effect of the Fermi momentum decreasing the reconstructed energy. 
	
%This discrepancy is caused by the higher relative importance of the Fermi momentum at  low neutrino energies. 

\begin{figure}
\includegraphics[width=8cm]{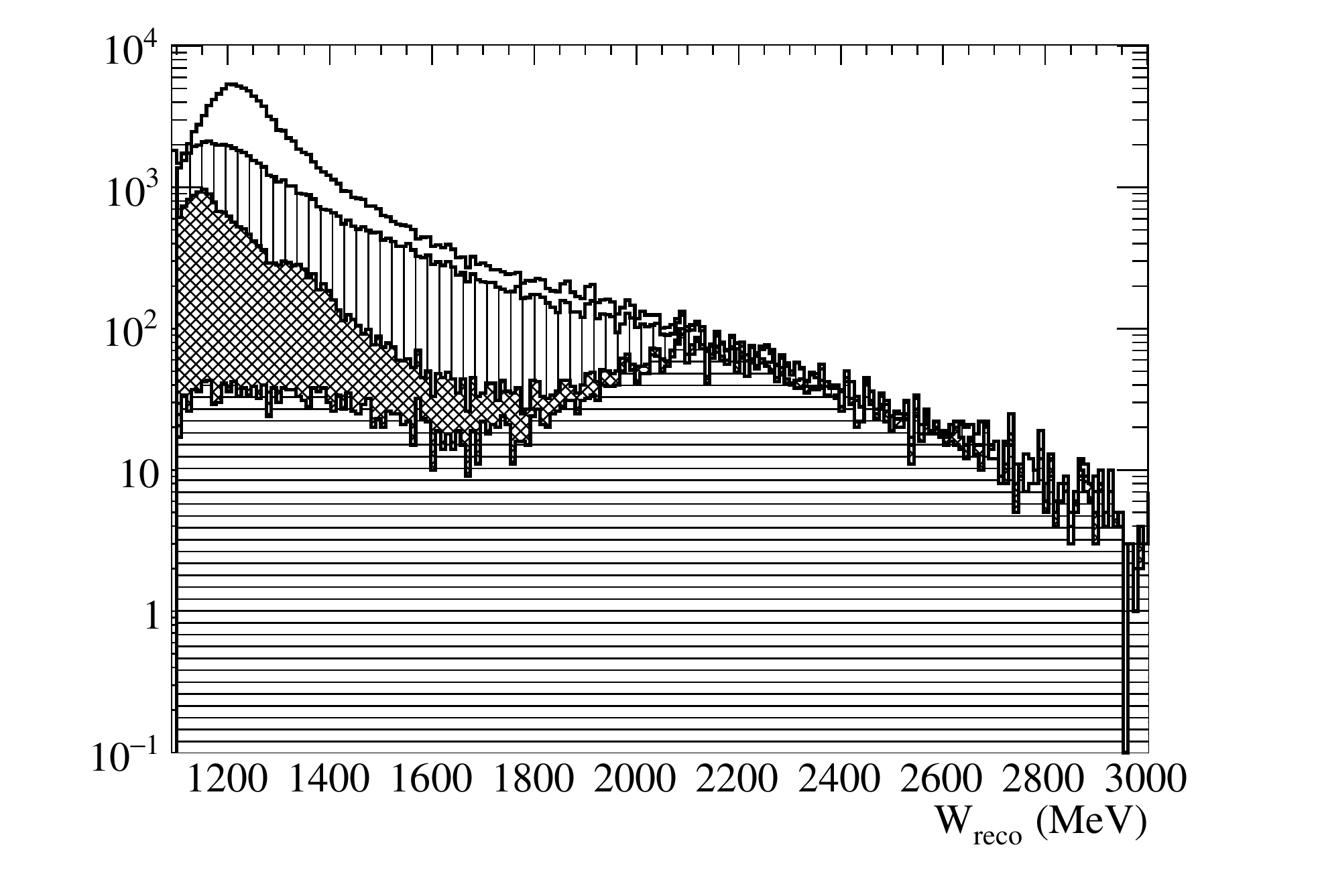}
\caption{\label{Fig:W}   Reconstructed  hadronic invariant mass. The empty histogram shows the CC one-$\pi$ events. The vertical line histogram shows the invariant mass when the pion momentum is modified by the FSI. The horizontal line histogram shows the CC-DIS contribution to the CC one-$\pi^+$ sample and the mesh-filled histogram shows the remaining backgrounds. }
\end{figure}

 A cut on the invariant mass of the p-$\pi^+$  system was suggested in the original paper \cite{Adler:1968tw}, and applied by the deuterium experiments \cite{ Barish:1978pj,Radecky:1981fn} to constrain the reactions into the region dominated by the $\Delta^{++}(1232)$ resonance.  Estimating the  neutrino  energy allows to obtain the invariant mass of the p-$\pi$ system, taken to be the $\mu-\nu$ invariant mass ($W_{reco}$), see Figure \ref{Fig:W}.  The plot shows the contribution of the different backgrounds; the deep inelastic background dominates at high invariant mass values. The invariant mass threshold ($W >$ 2 GeV) that defines the DIS region in NEUT is clearly seen in Figure \ref{Fig:W}. This seems to indicate that the reconstructed $W$ value is a sensitive observable in validating the implementation of the transition from multi-pion production to DIS in the Monte Carlo.

\begin{figure}
\includegraphics[width=8cm]{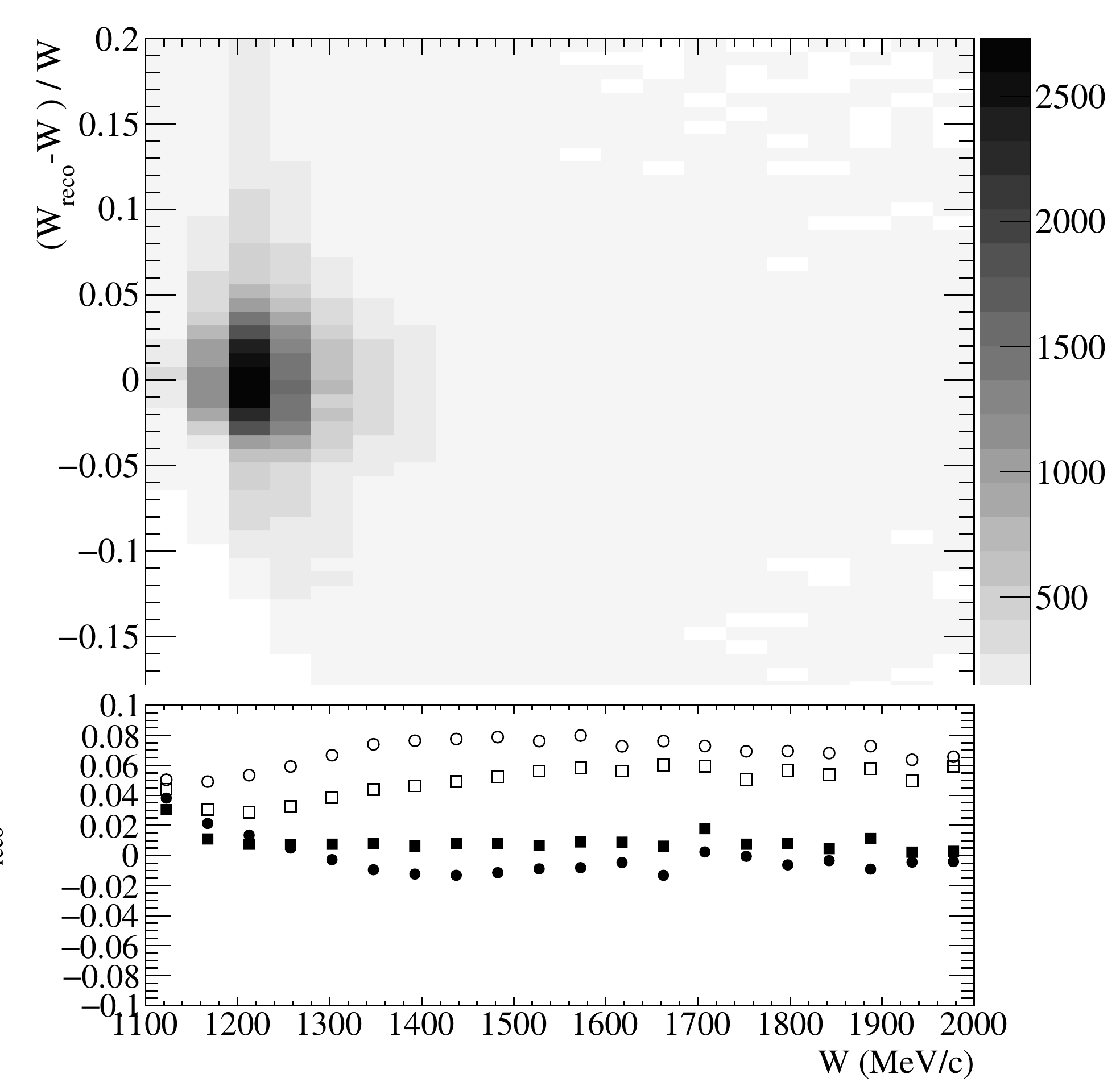}
\caption{\label{Fig:dWW}  Relative p-$\pi$  invariant mass reconstruction error ($ ( W_{reco} - W ) / W $ ) as a function of the true  p-$\pi$  invariant mass . The black circles (squares) represent the mean value of the relative error post(pre)-FSI. The empty circles (squares) show the RMS of the reconstructed invariant mass relative error post(pre)-FSI. }
\end{figure}

  The accuracy of the reconstruction is shown in Figure \ref{Fig:dWW}. The invariant mass reconstruction shows a small bias ( $<$ 4\%) over almost the full  range of W and also a relatively small RMS error ($\sim$ 8\%). The figure also shows the contribution of the Fermi momentum (black squares) and the combined effect of Fermi momentum and FSI (black circles).  The bias (below 1200~MeV) in the reconstructed $W$ is mainly caused by the Fermi momentum of the target nucleon. 
 
\subsection{ Reconstructed Adler angles}

\begin{figure}
\includegraphics[width=8cm]{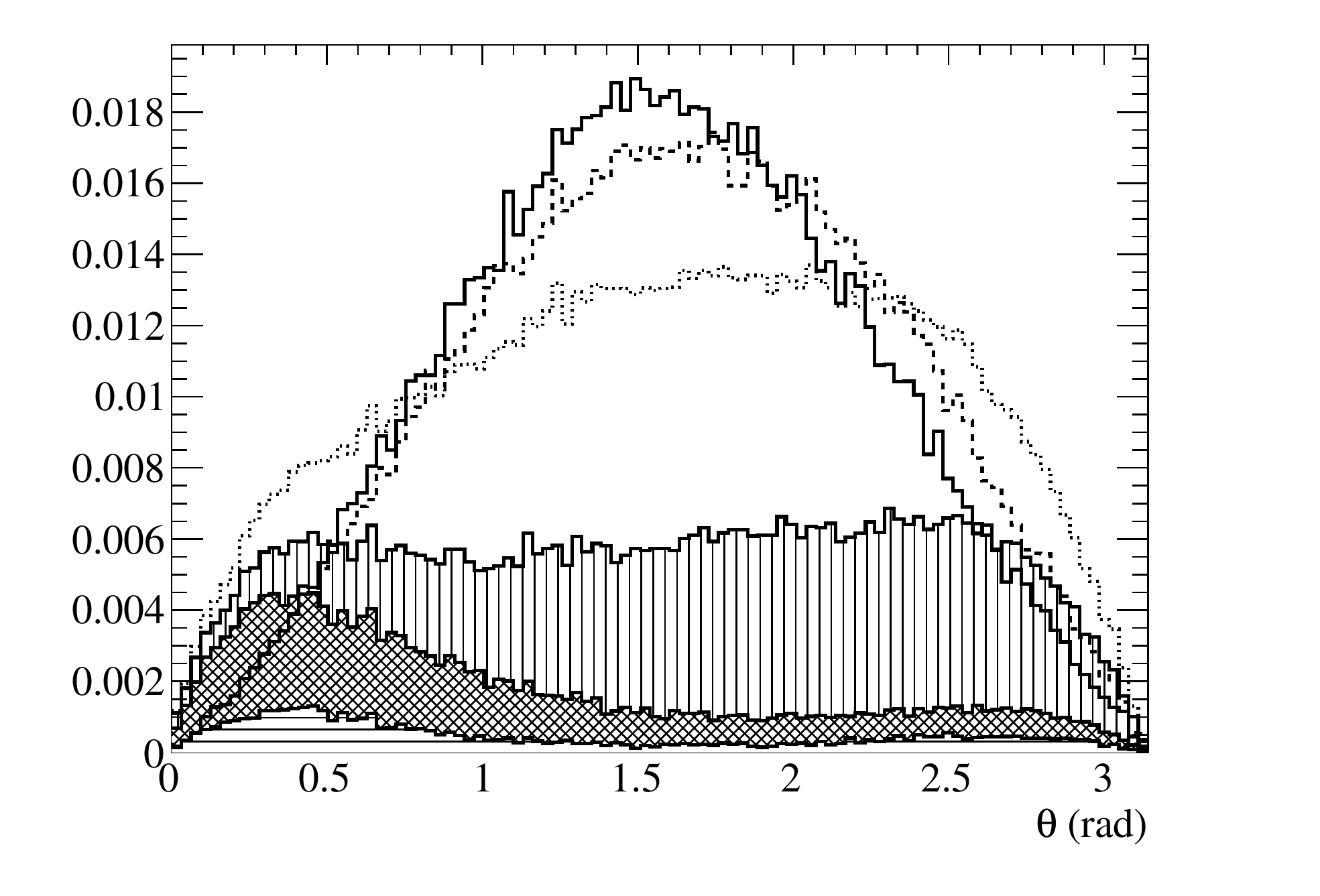}
\includegraphics[width=8cm]{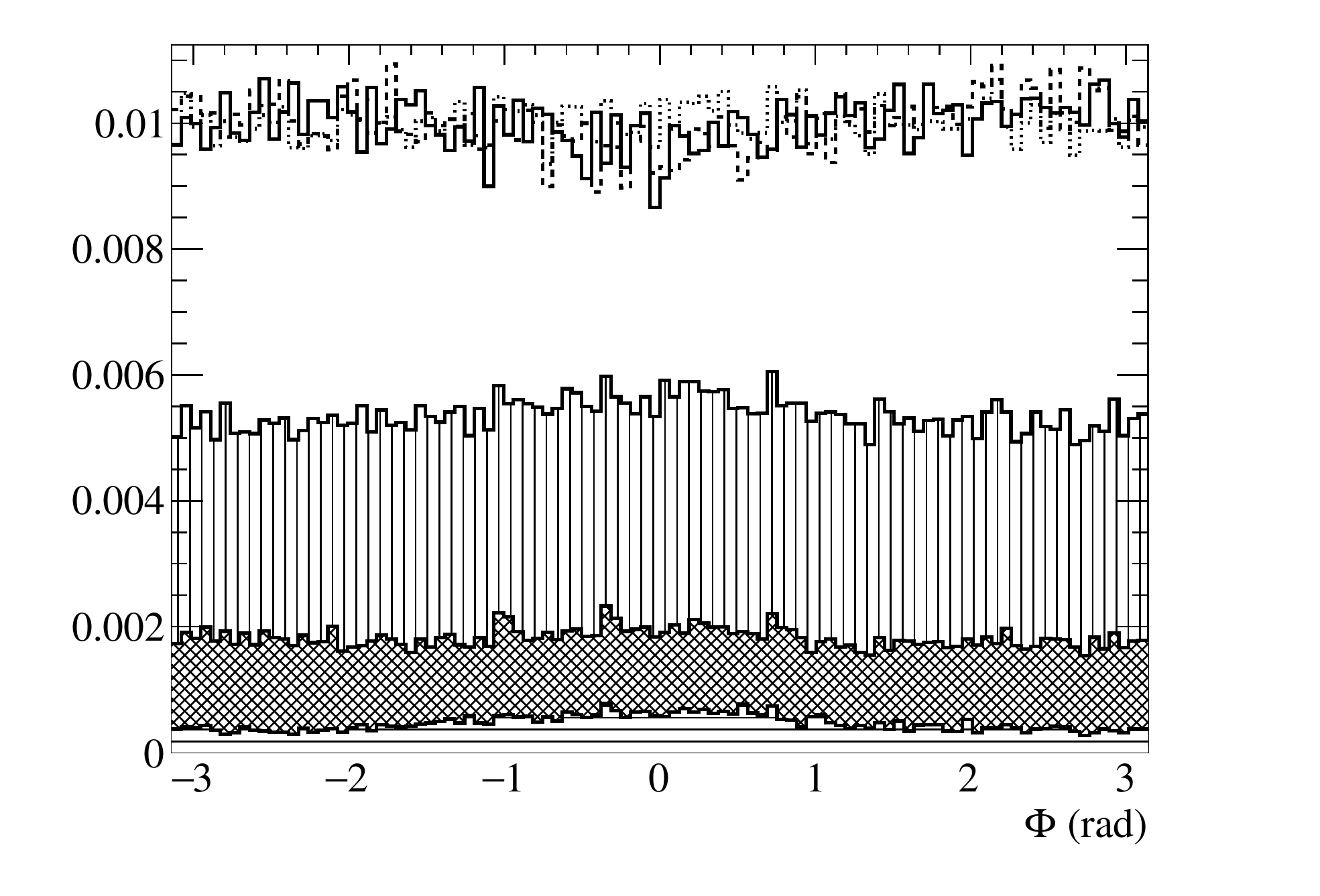}
\caption{\label{Fig:ThetaPhi}   True (solid line) and reconstructed (dotted line)  Adler $\theta$ angle (top panel) and $\phi$ angle (bottom panel). The true CC~1$\pi^+$ distributions include all interactions with the target nucleon and the reconstructed ones for those events with pions leaving the nucleus. The dashed line histogram shows the result when the Fermi momentum is ignored in the reconstruction of the true CC~1$\pi^+$ result. The histogram filled with vertical lines shows the CC~1$\pi^+$ events in which the pion momentum is~ modified by the FSI. The histogram filled with horizontal lines is the CC-DIS contribution to the CC~1$\pi^+$ sample and the histogram filled with a mesh contains the rest of the backgrounds. The distributions are normalized to unity. }
\end{figure}

 The reconstructed Adler angles are shown in Figure \ref{Fig:ThetaPhi}. The true distributions (solid line) are computed for all interactions at the nucleon level and the reconstructed distributions (dotted line) are shown for all  events with a single pion leaving the nucleus. 

% IS THE ENTIRE FOLLOWING PARAGRAPH INVISIBLE?
%The $\phi$ angle is almost unaffected by the pion traversing the nucleus, whereas the distributions of $\theta$ are different for the true and reconstructed %values due to the change of momentum of the pion while scattering within the nucleus but also due to the contribution of the backgrounds. On the contrary, the backgrounds are almost independent from the  $\phi$ angle.  
This first look at  the result allows to reach some preliminary conclusions: the transverse coordinates ($\phi$) are almost unaffected by nuclear effects while the longitudinal observable ($\theta$) is modified by the change in momentum of the outgoing pion. Background events tend to accumulate at low values of $\theta$ while $\pi^+$ rescattering events accumulate at high values of $\theta$ as it is shown in Figure \ref{Fig:ThetaPhi}

\begin{figure}
\includegraphics[width=8cm]{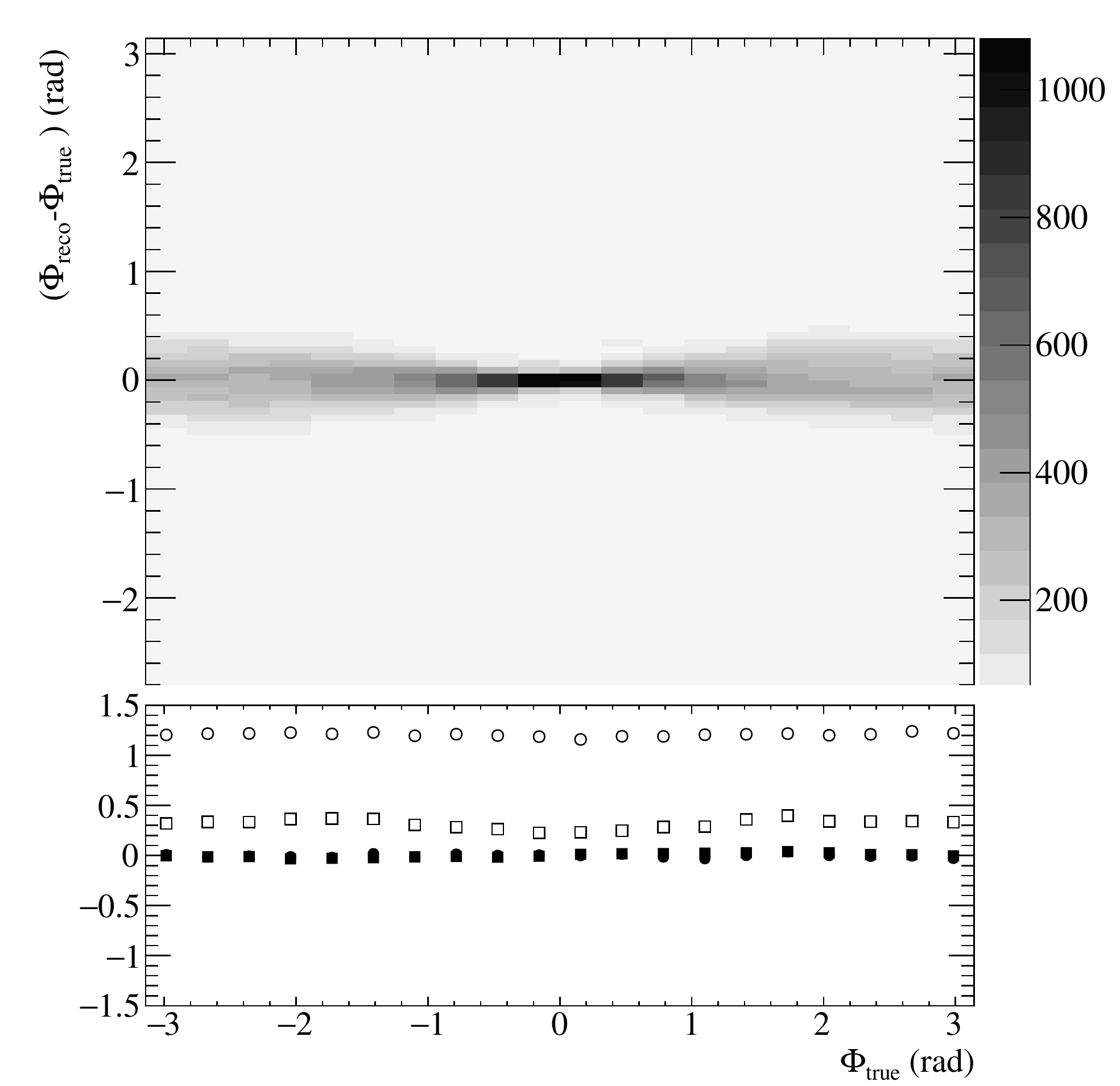}
\caption{\label{Fig:dPhivsPhi}  Reconstructed $\phi$ angle minus the true $\phi$ as a function of the $\phi$ angle. The black circles (squares) show the mean value of the difference post(pre)-FSI .   The empty circles(squares)  points show the RMS of the difference post(pre)-FSI.  }
\end{figure}

 The difference between the reconstructed and the true $\phi$ angle is shown in Figure \ref{Fig:dPhivsPhi}. The average bias is close to zero while the maximal RMS error is nearly 1.2~rad. The dependence on the true $\phi$ angle can be explained by the fact that angles around 0 and $\pi$~rad correspond to pions contained in the neutrino-muon reaction plane, see Figure \ref{Fig:AdlerReference}. These are the cases where the Fermi momentum that was ignored in the event reconstruction will produce the smallest  effect because the true motion of the target nucleon should be contained in the reaction plane. The comparison between the RMS due to the Fermi momentum (0.4~rad) and Fermi momentum plus FSI (1.2~rad), Figure \ref{Fig:dPhivsPhi}, indicates that the main contribution is the re-scattering of pion on its way out of the nucleus.  

\begin{figure}
\includegraphics[width=8cm]{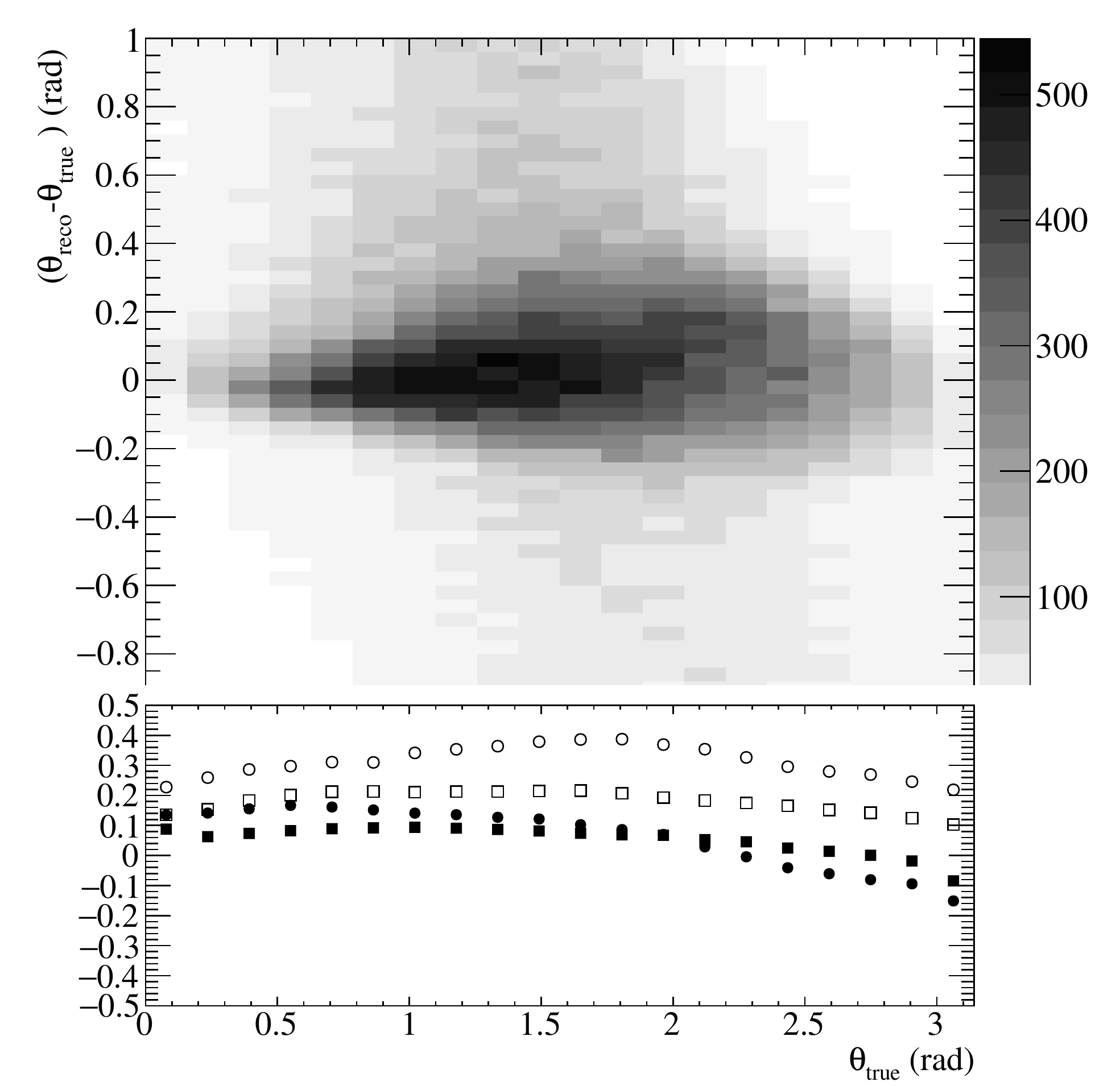}
\caption{\label{Fig:dThetavsTheta}  Reconstructed $\theta$ angle minus the true $\theta$ angle as a function of $\theta$. The black circles (squares) show the mean value of the difference post(pre)-FSI.  The empty circles(squares)  show the RMS of the difference post(pre)-FSI.  }.
\end{figure}

 The dispersion in the reconstructed values of $\theta$ is shown in Figure~\ref{Fig:dThetavsTheta}. In this case, the bias goes up to 0.15~rad and the RMS is as large as 0.4~rad. One can see that $\theta$ is very sensitive to the accuracy of the neutrino energy reconstruction and the intranuclear scattering of the charged pion. The FSI are the dominant contribution to the RMS of $\theta$ over its whole range.

\section{  Bias in asymmetry measurements }
	
\begin{figure}
\includegraphics[width=8cm]{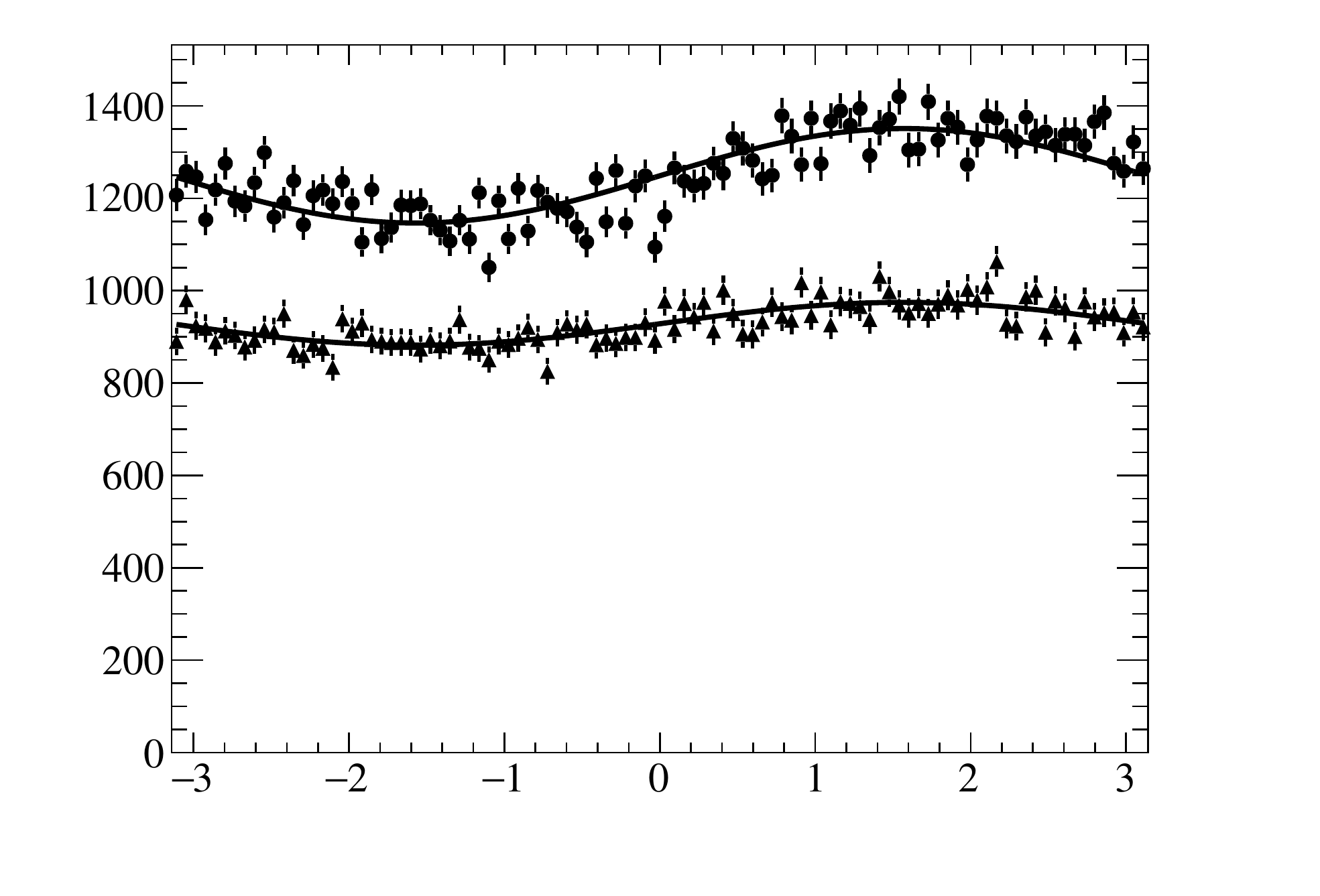}
\caption{\label{Fig:PhiFit}  $\phi$ distribution at the nucleon (black dots) and nucleus (black triangles) levels weighted by the angular dependence as described in the text. The solid line is the result of the fit to the function $A(1+\alpha\sin{\phi})$ }
\end{figure}

 We estimated the potential bias in the reconstruction of the Adler angles by means of a toy Monte Carlo. We take the dependence found from the deuterium results cited earlier \cite{Radecky:1981fn} and weigh the events according to the $\phi$ angle at the nucleon level with a simple parity-violating function:

$$ weight = ( 1 + \alpha \sin{ \phi } ) $$

where $\alpha$ is derived from the asymmetry calculation in the above paper \cite{Radecky:1981fn} to be 0.083. We fit the same angular dependence for the distributions of the $\phi$ angle at the nucleon and the nucleus level after subtracting the background and for events with W and $W_{reco}$ below 1400 MeV in order to select events dominated by $\Delta^{++}$ and $\Delta^{+}$ resonant contributions; see Figure~\ref{Fig:W}. The results are shown in Figure \ref{Fig:PhiFit}. The fitted $\alpha$ values, $0.082\pm0.004$ at the nucleon level and $0.053\pm0.005$ at the nucleus level are similar. The errors shown only include the effects of the Monte Carlo statistics. For the reconstructed $\phi$, we have subtracted the non-CC~1$\pi^+$ interactions. Background modeling will introduce an additional source of uncertainty that is not evaluated in this study.

The bias in the determination of the asymmetry was estimated by generating several $\phi$ functional dependencies and adjusting the angular dependence. The results are shown in Table \ref{Table:Alpha}. Although there is a bias in the determination of the $\alpha$ parameter caused by the smearing of the reconstructed $\phi$ angle, there is still sensitivity to the determination of $\alpha$. An accurate experimental measurement should consider the variation of the polarization with the event kinematics \cite{Hernandez:2007qq}. 
%One important caveat is the fact that the main contribution to the $\phi$ reconstruction uncertainty is the model of the nucleon target at rest.  The reconstructed %bias shown in Table \ref{Table:Alpha} is related to the nucleon Fermi momentum and nuclear model. 
Table~\ref{Table:Alpha} shows the result when we take into account the effect of the Fermi momentum.  The effect of the Fermi momentum seems to be negligible at this level, as expected from Figure~\ref{Fig:dPhivsPhi}. The same calculation was performed for an angular dependence of the type $(1+\alpha \sin{2\phi})$. The results are shown in Table~\ref{Table:Alphasin2} . The results are worse than in the previous case due to the convolution with the reconstructed $\phi$ RMS of the faster oscillation frequency. 

\begin{table*}
\caption{\label{Table:Alpha} The values of the angular dependence fits ($\alpha$) for different values of transverse polarization and $\sin{\phi}$ dependence. Results are shown for pions before and after the FSI. }
\begin{ruledtabular}
\begin{tabular}{lcrcr}
 & \multicolumn{2}{c} { post-FSI } & \multicolumn{2}{c} {pre-FSI} \\
\hline
$\alpha_{true}$  & $\alpha_{reco}$  & $ {\alpha_{true}-\alpha_{reco} \over \alpha_{true} }$ & $\alpha_{reco}$   &  $ {\alpha_{true}-\alpha_{reco} \over \alpha_{true} }$  \\
\hline

0.02 & 0.0166 $\pm$  0.0050 & -0.17  &  0.0248 $\pm$  0.0055 & 0.243 \\ 
0.04 & 0.0285 $\pm$  0.0050 & -0.29  &  0.0440 $\pm$  0.0055 & 0.101 \\ 
0.06 & 0.0404 $\pm$  0.0050 & -0.32  &  0.0632 $\pm$  0.0055 & 0.054 \\ 
%0.08 & 0.0524 $\pm$  0.0050 & -0.34  &  0.0823 $\pm$  0.0055 & 0.030 \\ 
0.10 & 0.0644 $\pm$  0.0050  & -0.36  &  0.1015 $\pm$  0.0054 & 0.015 \\ 
0.12 & 0.0763 $\pm$  0.0050 & -0.36  &  0.1206 $\pm$  0.0054 & 0.005 \\ 
%0.14 & 0.0882 $\pm$  0.0050 & -0.37  &  0.1400 $\pm$  0.0054 & -0.001 \\ 
%0.16 & 0.1001 $\pm$  0.0050 & -0.37  &  0.1589 $\pm$  0.0054 & -0.007 \\ 
0.18 & 0.1121 $\pm$  0.0050 & -0.38  &  0.1781 $\pm$  0.0053 & -0.011 \\

\end{tabular}
\end{ruledtabular}
\end{table*}

\begin{table*}
\caption{\label{Table:Alphasin2} The values of the angular dependence fits ($\alpha$) for different values of transverse polarization and $\sin{2\phi}$ dependence. Results are shown for pions pre- and post-FSI. }
\begin{ruledtabular}
\begin{tabular}{lcrcr}
 & \multicolumn{2}{c} { post-FSI } & \multicolumn{2}{c} {pre-FSI} \\
\hline
$\alpha_{true}$  & $\alpha_{reco}$  & $ {\alpha_{true}-\alpha_{reco} \over \alpha_{true} }$ & $\alpha_{reco}$   &  $ {\alpha_{true}-\alpha_{reco} \over \alpha_{true} }$  \\
\hline
0.02 & 0.0078 $\pm$  0.0050 & -0.60  &  0.0129 $\pm$  0.0055 & -0.35 \\ 
0.04 & 0.0192 $\pm$  0.0050 & -0.52  &  0.0308 $\pm$  0.0055 & -0.23 \\ 
0.06 & 0.0305 $\pm$  0.0050 & -0.49  &  0.0487 $\pm$  0.0055 & -0.19 \\ 
%0.08 & 0.0418 $\pm$  0.0050 & -0.48  &  0.0666 $\pm$  0.0054 & -0.17 \\ 
0.10 & 0.0531 $\pm$  0.0050 & -0.47  &  0.0845 $\pm$  0.0054 & -0.16 \\ 
0.12 & 0.0644 $\pm$  0.0050 & -0.46  &  0.1024 $\pm$  0.0054 & -0.15 \\ 
%0.14 & 0.0757 $\pm$  0.0050 & -0.46  &  0.1203 $\pm$  0.0054 & -0.14  \\ 
%0.16 & 0.0870 $\pm$  0.0050 & -0.46  &  0.1382 $\pm$  0.0054 & -0.14 \\ 
0.18 & 0.0983 $\pm$  0.0050 & -0.45  &  0.1561 $\pm$  0.0054 & -0.13   \\
\end{tabular}
\end{ruledtabular}
\end{table*}

We have also estimated the bias in the forward-backward asymmetry. Events are not reweighted and the asymmetry is computed as: 

$$ A_{FB} = { N_{\cos{\theta}>0} -  N_{\cos{\theta}<0} \over  N_{\cos{\theta}>0} + N_{\cos{\theta}<0} } $$
for the distributions of $\theta$ both at the nucleon and the nucleus levels, after removing the background. The values of the resulting asymmetries are: $-0.007\pm 0.003$ (as predicted by the NEUT Monte Carlo) and $-0.179\pm0.003$ from the reconstructed observable. The observed bias is produced by the FSI and Fermi momentum within the nucleus because the Adler $\theta$ is very significantly modified, as shown in  Figure~\ref{Fig:ThetaPhi}. The dependence of the $\theta$ angle on the FSI and Fermi momentum makes it a very useful observable when  investigating the nuclear effects on the results of the reaction.

\section{ Conclusions }

We have shown that it is possible to measure the Adler angles in neutrino-nucleus scattering. The results based on the NEUT Monte Carlo show that one can determine the transverse polarization of the $\Delta$ resonance because the information is reasonably well maintained despite the FSI and the need to reconstruct the energy of the incoming neutrino from the experimental data. The longitudinal polarization is shown to depend strongly on the kinematics of the emerging pion, but it appears to allow investigatingf the effects of pion re-scattering, high mass resonances and deep inelastic processes on the CC one-$\pi^+$ tracks emerging from the nucleus. The results indicate that current high-statistics experiments can explore complex observables like Adler angle as a function of the kinematic parameters of the scattering process such as the  energy of the neutrino, the hadronic invariant mass and four-momentum transfer.

\begin{acknowledgments}

The author acknowledges the support received from the Ministerio de Economia y Competitividad under grants FPA2014-59855 and Centro de Excelencia Severo Ochoa SEV-2012-0234, some of which include FEDER funds from the European Union. He would like to to acknowledge support from Y. Hayato on the NEUT Monte Carlo, discussions with J. Nieves on the implications of Adler angle measurements and help from S. Bordoni and M. Cavalli-Sforza in revising this paper.

\end{acknowledgments}


\bibliography{Bibliography.bib}

\end{document}